\documentclass[]{spie}  %>>> use for US letter paper
\usepackage[]{graphicx}
\usepackage{aas_macros}

\title{Monte-Carlo Imaging for Optical Interferometry} 

\author{Michael~J. Ireland\supit{a}, John D. Monnier\supit{b} and
Nathalie Thureau\supit{b}
\skiplinehalf
\supit{a}Caltech, MC 150-21,1200 E. California Blvd., Pasadena, CA 91125, USA;\\
\supit{b}Department of Astronomy, University of Michigan, 501 East University
Av., Ann Arbor, \\ MI 48109, USA}

\authorinfo{Further author information: E-mail: mireland@gps.caltech.edu}

\begin{document} 
  \maketitle 

%%%%%%%%%%%%%%%%%%%%%%%%%%%%%%%%%%%%%%%%%%%%%%%%%%%%%%%%%%%%% 
\begin{abstract}
We present a flexible code created for imaging from the
bispectrum and $V^2$. By using a simulated annealing method, we limit the
probability of converging to local chi-squared minima as can occur
when traditional imaging methods are used on data sets with limited
phase information. We present the results of our code used on a simulated 
data set utilizing a number of
regularization schemes including maximum entropy. Using the
statistical properties from Monte-Carlo Markov chains of images, we
show how this code can place statistical limits on image features such
as unseen binary companions.
\end{abstract}

%>>>> Include a list of keywords after the abstract 


%%%%%%%%%%%%%%%%%%%%%%%%%%%%%%%%%%%%%%%%%%%%%%%%%%%%%%%%%%%%%
\section{INTRODUCTION}
\label{sect:intro}  

It is well known that a large class of images can be consistent with a
particular interferometric data set. This is more true for optical
inferferometry than radio interferometry, due to the general
unavailability of absolute visibility phase. An imaging algorithm such
as CLEAN or Maximum Entropy combined with self-calibration attempts to find
the `best' possible image consistent with the interferometric data. Both finding this
`best' image and interpretation of features within the image can be
difficult, and in general requires some kind of
regularization. Regularization punishes images that look 'bad' (such
as having too much unresolved structure) to find a compromise between
lowering the $\chi^2$ statistic and achieving an optimal regularization statistic.

The imaging code MACIM described in this paper is a Monte-Carlo
Markov chain algorithm that aims to both reliably find the global
minimum of a regularized $\chi^2$ statistic in image space, and to
characterize this minimum. The algorithm can operate without any
regularization to find images that are optimal in the Bayesian
sense. In this mode, the code can also characterize the joint
probability density of images consistent with the data. Alternatively,
the code can combine model-fitting and imaging or use novel
regularizations based on {\em a priori} imaging constraints such as the
expected existence of connected regions of zero flux.

\subsection{Markov Chains and Bayesian Inference}

Bayes theorem states that the probability that a model $\theta$ (i.e. an image in
our context) is correct given a given data set D (which includes errors on the
data) is \cite{Gamerman97}:

\begin{equation}
 p(\theta|D) = \frac{f(D|\theta)p(\theta)}{f(D)},
 \label{eqnBayes}
\end{equation}

where

\begin{equation}
 f(D) = \int f(D|\theta)p(\theta) d\theta.
 \label{eqnBigIntegral}
\end{equation}

Here $p(\theta)$ is the prior distribution of $\theta$, and
$p(\theta|D)$ is the posterior distribution. In the case of
independent Gaussian errors, the likelihood function
$f(D|\theta)$ takes a multivariate Gaussian form:

\begin{equation}
 f(D|\theta) \propto exp(\sum (D_m(\theta) - D)/2\sigma_i^2) = exp(\chi^2/2)
\end{equation}

Here $D_m(\theta)$ is the model data ($V^2$, bispectrum) corresponding to the image $\theta$.
For the context of imaging, a regularization technique is contained in the pre-determined prior
distribution $p(\theta)$.

When imaging or tackling many other problems with high dimensionality, the integral in
Equation~\ref{eqnBigIntegral} can not be evaluated in a reasonable
time, so it is not possible to explicitly evaluate
Equation~\ref{eqnBayes}. An alternative to explicit evaluation is to use a Monte-Carlo
Markov Chain technique to sample the regions of image space where
$p(\theta|D)$ is highest \cite{Gamerman97}. The distribution of images in the resultant
Markov Chain $\theta_j$ then becomes a discrete version of the
posterior distribution $p(\theta|D)$ from which inferences on the set
of possible images can be made.

\section{{\em MACIM} IMAGING ALGORITHM}

\subsection{General Algorithm}

The general algorithm used by MACIM is a simulated annealing
algorithm with the Metropolis sampler \cite{Bremaud99,Gamerman97}. The image
state space $\theta_j$ at iteration $j$
consists of the set of pixel vectors $\{p_i\}_j$ 
for all flux elements $i$ with $1 \leq i \leq \lambda$. $\lambda$
is the total number of flux elements. The flux in the image is
constrained to be equal to 1, unless model fitting of
Section~\ref{sectModel} is used. The vectors $p_i$ exist on a
finite square grid with resolution at least $\lambda/4\max(\{B\})$, with
$\max(\{B\})$ the maximum baseline length. There are two classes of steps
that the algorithm can take. The first class of step moves a flux
element, i.e. it randomly chooses a flux element $I$, modifies $p_I$
to form the {\em tentative} state 
$\{q_i\} =  \{p_1,p_2,...,p_I+s,...,p_n\}$ for some step $s$, chosen to be in a
  random direction. Given a temperature $T$, the modification to the image state
is accepted with a probability:

\begin{equation}
 p(j,j+1) = \min(1,\exp(\frac{\chi^2(\{p_i\}_j) - \chi^2(\{q_i\})}{2T} + \alpha \Delta R)).
 \label{eqnPMovement}
\end{equation}

Here  the $\chi^2$ function is the total $\chi^2$ function calculated
directly from the interferometric data in {\tt oifits} format ($V^2$,
bispectrum, complex visibility). $\Delta R$ is the change in the
regularization parameter $R$ and $\alpha$ is a regularization scaling
parameter. If the tentative state is accepted, then 
$\{p_i\}_{j+1}$ is set to $\{q_i\}$. Otherwise, we set
$\{p_i\}_{j+1} = \{p_i\}_j$. 

The tentative moves for $p_i$ include several different types of
flux steps $s$: moving one or several pixels along one of the image axes,
moving the flux unit anywhere in the image, or moving to the location
of another randomly selected flux element. Large steps in general have a smaller
probability of success than small steps. For this reason, the step
type for the tentative move is chosen so that on average the
probability of accepting the tentative state is between 0.2 and
0.45. For all steps $s$, the probability of choosing the tentative reverse
transition at random is equal to the probability of choosing the
forward transition.

The configuration space entropy (i.e. the logarithm of the image
degeneracy) does not explicitly come into
Equation~\ref{eqnPMovement}, but does enter the picture if one wishes
to find the most probable, or mode image. To understand why this is,
consider the image representation $\{N_k\}$ where $N_k$ represents the
number of flux elements in pixel $k$. Two images with the
same $\{N_k\}$ are equal but can be degenerate, as each can be formed by
a number of possible state vectors $\{p_i\}$. We will follow the
notation of Ref.~\citenum{Sutton06}, and call this the multiplicity
$W$:

\begin{equation}
 W = \frac{\lambda !}
          {N_1 ! N_2 !...N_n !}.
 \label{eqnWDefn}
\end{equation}

Here $n$ is the total number of pixels in the image. Changing the
total number of image elements $\lambda$ was done in
Ref.~\citenum{Sutton06} by assuming a uniform
prior on $\lambda$: all numbers of non-zero flux elements were assumed
equally probable. This means that
the normalized prior distribution of $\{N_k\}$ is given by:

\begin{equation}
 p(\{N_k\}) = \frac{W}{n^{(1-\delta) \lambda}},
 \label{eqnPriorNk}
\end{equation}

with the parameter $\delta = 0$. In general, this prior distribution
for $\{N_k\}$ with $\delta = 0$ does not give `sensible' images when
$\lambda$ was permitted to vary (the Markov chain would converge at
very high $\chi^2$ and a low number of elements). Therefore, non-zero
values of, $\delta$ can be input as an optional parameter. Note that
$\delta=1$ is equivalent to the prior distribution for $p(\{N_k\})$
that all configurations are equally probable, as opposed to all values
of $\lambda$ being equally probable.

The second class of step consists of adding a new flux element in
pixel $K$ or removing flux element $I$ (i.e. changing $\lambda$). This step is intrinsically
asymmetrical, as the probability of the reverse step not equal to the
forward step. However, for $\delta=0$, the ratio of the probabilities of the forward and
reverse steps is equal to the inverse of the ratio of the prior
probability from Equation~\ref{eqnPriorNk}, meaning that
Equation~\ref{eqnPMovement} is still appropriate for determining the
Metropolis algorithm acceptance probability. For other values of
$\delta$, the exponential function in Equation~\ref{eqnPMovement} is
multiplied by $n^{\delta}$ when adding a flux element, and
divided by $n^{\delta}$ when removing a flux element. For removing flux element
$I$, we use $\{q_i\} =  \{p_1,p_2,...,p_{I-1},p_{I+1},...,p_n\}$, and
for adding to pixel $K$ we use $\{q_i\} = \{p_i,K\}$.

The annealing temperature $T$ is modified based on the reduced
$\chi^2$, $\chi^2_r$, according to the following algorithm:

\begin{equation}
 T_{j+1} = T_j + \frac{(\chi^2_{r,j} - \gamma T_j)(1 - \chi^2_t/\chi^2_{r,j})}
   {\Delta j }
\end{equation}

The parameter $\gamma$ is always greater than 1 (set to 4 by
default). The other parameters are the reduced $\chi^2$ target
$\chi^2_t$ and the timescale of temperature changes $\Delta j$. This
algorithm fixes $T_j$ to be near $\chi^2_{r,j}/\gamma$ during
convergence, and then fixes $\chi^2_r$ to be near $\chi^2_t$ once the
algorithm has converged. A minimum temperature limit $T_{\min}$ can also be
placed on $T_j$.

\subsection{Regularizers}

There are two currently implemented regularizers in MACIM, although
many are possible given that no derivative is required, as is often
the case for other imaging algorithms. The first
regularizer is simply the maximum entropy regularizer $R =
\log(W)$. With a sufficient number of flux elements, MACIM can
therefore be used to find the maximum entropy regularized image. The
second implemented regularizer is a dark interaction energy
regularizer. This regularizer is the sum of all pixel boundaries with
zero flux on either side of the pixel boundary. Inspired by the Ising
model, this regularizer encourages large regions of dark space
in-between regions of flux and represents a means to utilize {\em a priori}
knowledge of source structure.

\subsection{Model Fitting}
\label{sectModel}

For certain astrophysical targets, a combination of model-fitting and
imaging can significantly aid in data interpretation. An example of
this is the point-source plus extended flux images of VY~Cma and
NML~Cyg in Ref.~\citenum{Monnier04} where the use of a maximum entropy
prior with a central point source changed the image morphology
significantly. Model fitting is combined with imaging in MACIM by
varying model parameters simultaneously with flux movement. Currently,
the only implemented model fitting option is a centrally-located 
uniform disk (or point source) that takes up some fraction of the total
image flux, and an over-resolved (background) flux component. This
model has three parameters: flux fraction of the central source,
diameter of the central source and over-resolved flux fraction. Any of
these parameters can be fixed or be allowed to move freely according
to the Metropolis-Hastings algorithm. The parameter step sizes are
chosen so that the probability of accepting the tentative new
parameter is on average 0.3.

\subsection{Specific Modes of Operation}

There are several ways in which MACIM can create images:

\begin{itemize}
\item Bayesian mean map. This is the default mode of operation, with
  the settings $T_{\min}=1$, $\chi^2_t = 0$, $\lambda$ fixed at the
  number of input degrees of freedom and no regularization. Starting
  from an initial map (by default a 
  point source), the simulated annealing algorithm converges to a
  global minimum where as long as $\chi^2_r < \gamma$ (default
  $\gamma=4$) we have  $T=1$. Once we have $T=1$, the properties of
  Markov chains enable the full posterior distribution of images to be
  sampled. Optionally, the full chain can be output instead of just
  the Bayesian mean. 
\item Variable-$\lambda$ Bayesian mean map. By choosing $\lambda_{\rm
  min} < \lambda_{\rm max}$, the number of image elements is permitted
  to vary. Due to frequent convergence problems, $\delta$ is set to 0.1 rather
  than 0 by default.
\item Bayesian mode map. This output is also output whenever the
  Bayesian mean map is output. The average of a number of images near the maximum of
  $p(\{N_k\})$ can be output, giving a variant of a maximum-entropy
  map (a maximum multiplicity map). Due to the quantization of flux,
  averaging a number of images (say, 10\% of the final chain) about
  the mode is more aesthetically pleasing than the the single mode
  map. An alternative to this kind of averaging is using the single
  mode map to adaptively bin the image plane according to the level of
  quantization noise.
\item Pseudo-maximum entropy map. By setting $T_{\min}=0$, $\chi^2_t = 1$ and
  fixing $\lambda$ to a large number (e.g. double the number of input
  degrees of freedom), multiplicity (which converges to entropy for
  large $\lambda$) is maximized while fixing $\chi^2_r=1$. Three kinds
  of potentially useful maps are simultaneously output in this case:
  the mean map, the mode map and the `maximum entropy' map, which
  contains the same number of images  as the mode map but weights the
  multiplicity so that the mean $\chi^2_r$ in the final image is 1. 
\item Regularized map. In general, model fitting and dark interaction
  energy regularization is most easily performed using a fixed
  $\lambda$. Clearly a large range of possible input parameters are
  possible here, depending on the exact nature of any {\em a priori}
  information. 
\end{itemize}

\subsection{Difficulties and Future Work}

The seemingly greatest difficulty in using MACIM to make images is
choosing the value of $\lambda$ (or $\delta$ if $\lambda$ is allowed
to vary). One argument for an optimal $\lambda$ choice comes from the
requirement that MACIM converges and well-samples the posterior
distribution.

The optimal value of the acceptance probability $p(j,j+1)$ is thought
to be in the range of 0.2 to 0.5 \cite{Gamerman97}. For acceptance
probabilities outside this range, the Markov Chain samples the
posterior distribution at a much slower rate. Acceptance probabilities
in this range can only be found at moderate values of $\lambda$. For
this reason, MACIM can not well sample the posterior distribution in the
high-$\lambda$ limit (where it becomes just like for the MEM
algorithm) or in the low-$\lambda$ limit (as occurs if $\delta$ is set
near zero). 

Another argument for optimal $\lambda$ may be the desire that
the mean value of $\chi^2_r$ is 1.0. In principle, there is a minimum
in mean $\chi^2_r$ (generally less than 1.0) at some value of $\lambda
= \lambda_{\min}$, and $\chi^2_r$ increases on either side of this
minimum\cite{Sutton06}. Therefore, there should be two $\lambda$
values for which $\chi^2_r=1$. However, the lower-$\lambda$ value for
mean $\chi^2_r = 1$ can not be well sampled, because of the very high
barriers to flux movement or adding/removing flux elements. Certainly
this problem will require more work for either completely automatic operation
of MACIM or at least a well-defined knowledge of the influence of
$\lambda$ on deriving statistical inferences from the output MACIM
Markov Chain.

\section{SOFTWARE IMPLEMENTATION}

MACIM is written in the {\tt c} programming language, with an option
for multi-threaded operation (multiple Markov chains running
simultaneously, which are combined on completion). The transform
between image-space and complex visibility is stored in memory as
vectors containing $\exp(i u_m x_k)$ and $\exp(i v_m y_k)$ for baselines
$m$ and pixels $k$. $u_m$ and $v_m$ are the standard $u$ and $v$
coordinates for baseline $m$. Splitting the pixel coordinates into
$x_k$ and $y_k$ in this fashion means that only $2 M_b \sqrt{n}$
complex numbers need to be stored in memory (with $M_b$ the number of
baselines). At each iteration, the mathematical functions required are
limited to elementary arithmetic operations and one evaluation of the
exponential function (Equation~\ref{eqnPMovement}). No evaluations of
FFTs, trigonometric functions or square roots are required. For this reason,
the millions of iterations required to characterize the posterior
distribution can be run on a 2\,GHz class computer in several minutes
for a typical modern interferometric data set.

Only one argument is required to run MACIM: the input {\tt
oifits} file name. However, the default image size of
$\lambda/\min(B)$ with $\min(B)$ the minimum baseline length is
often not appropriate for a given data set. The maximum number of image
elements $\lambda_{\max}$ and the pixel scale are other parameters
that sometimes should be tweaked for optimal performance. Furthermore,
if convergence is too time consuming, then it can be advantageous to
run MACIM with a fixed small $\lambda$ until convergence, and then
increase $\lambda_{\max}$ with the converged image as an initial model
{\tt fits} file.

\begin{figure}
 \includegraphics[scale=1.05]{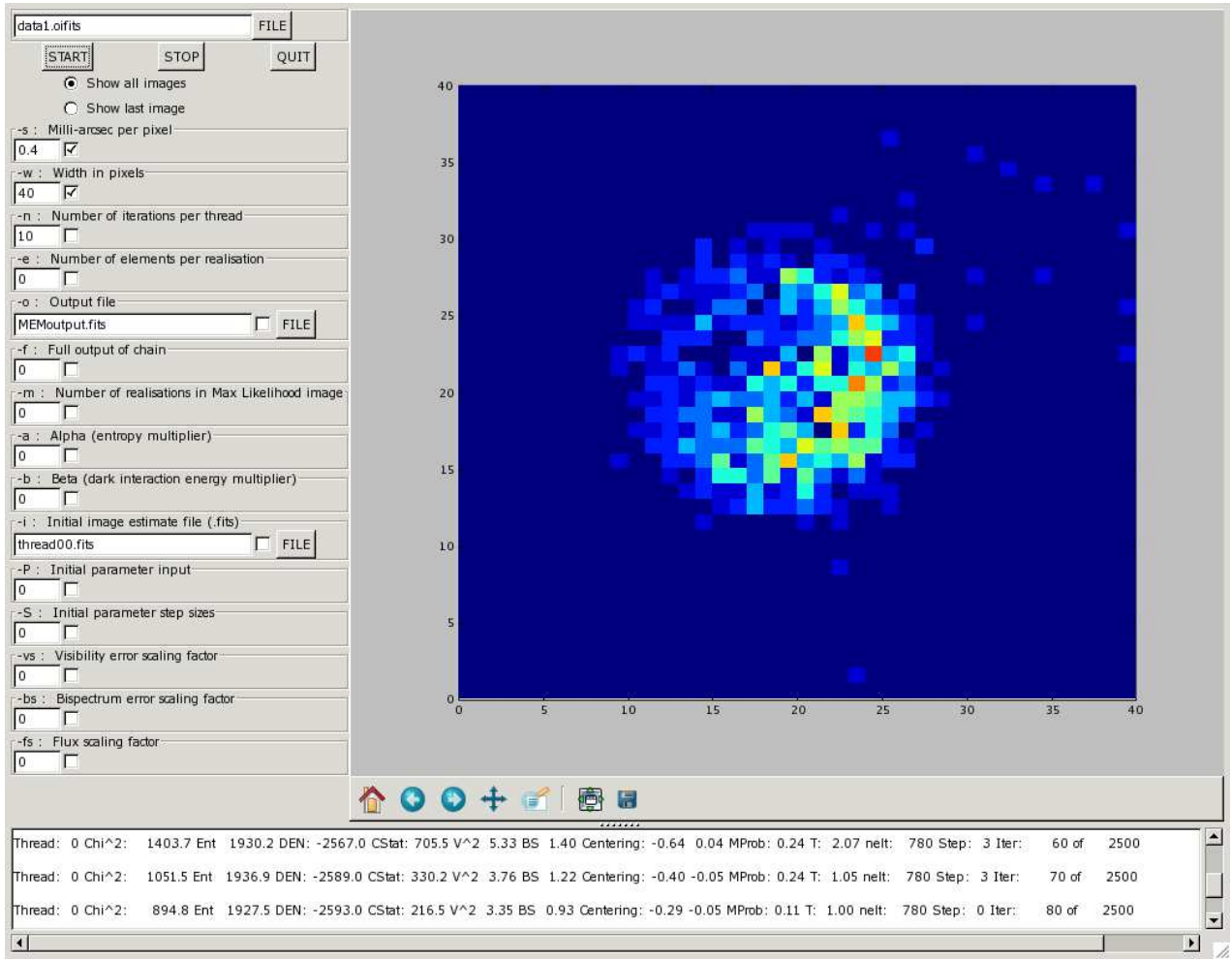}
 \caption{A screen shot of the MACIM GUI, showing a single image from
 the Markov Chain at a point where the algorithm has almost
 converged. The parameters in the lower window include the   
 algorithm temperature $T$, the $\chi^2_r$ values for $V^2$ (V\^2) and bispectrum (BS).}
 \label{figGui}
\end{figure}

MACIM is run via the {\tt unix} command line or a {\tt
python}-based graphical user interface (GUI). A screen-shot of the GUI
is shown in Figure~\ref{figGui}, demonstrating the display of an image
$\{N_k\}_j$, the output from MACIM and text boxes for easy optional
parameter input. There are also {\tt IDL}
tools available in the MACIM distribution for graphically displaying the images
$\{N_k\}_j$ and the $\chi^2_r$ values as MACIM converges and explores
the global minimum. MACIM has been successfully tested on solaris,
linux and Mac OS X platforms.

\section{EXAMPLES OF ALGORITHM PERFORMANCE}

\begin{figure}
  \begin{minipage}{7.0cm}
    \hspace{0.3cm}
    \includegraphics[scale=0.7]{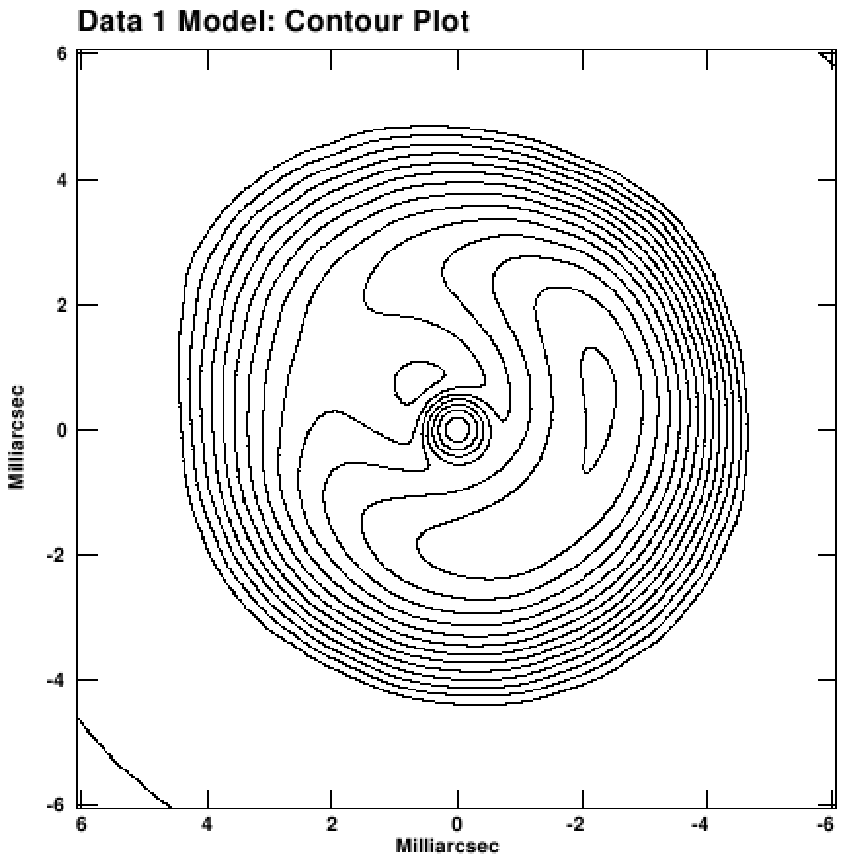}
  \end{minipage}
  \begin{minipage}{7.0cm}
    \hspace{0.3cm}
    \includegraphics[scale=0.75]{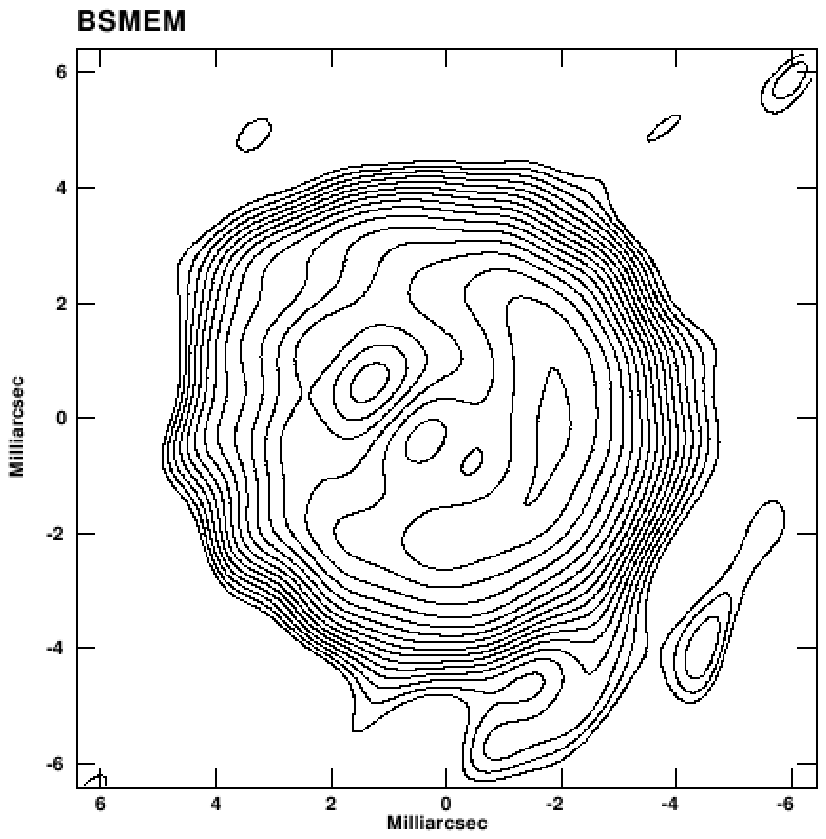}
  \end{minipage}
  \begin{minipage}{7.0cm}
    \includegraphics[scale=0.75]{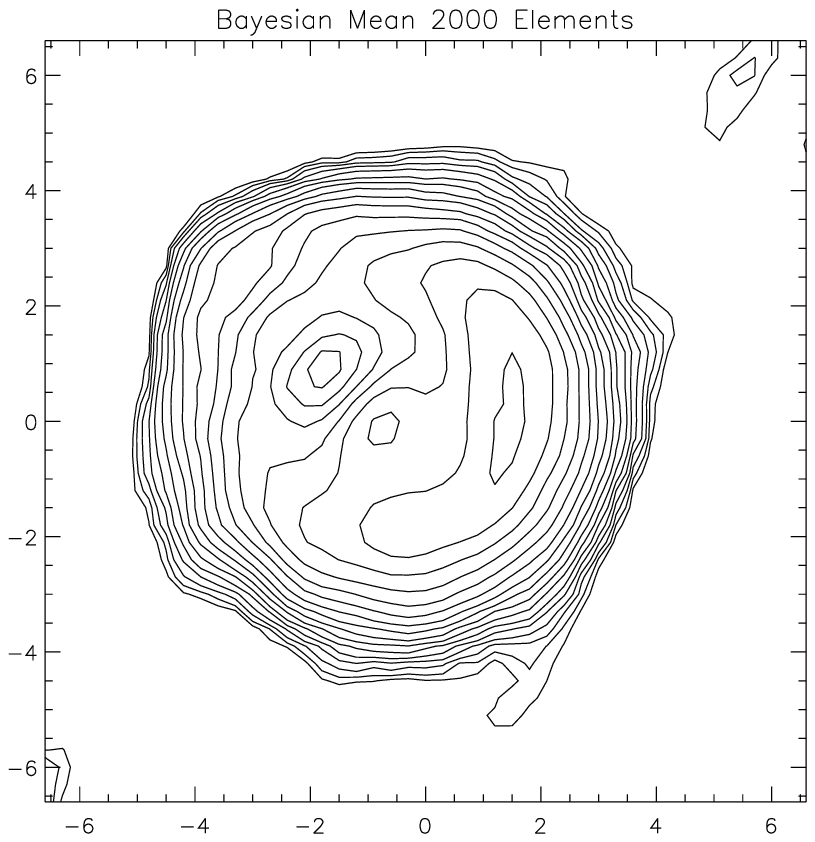}
  \end{minipage}
  \begin{minipage}{7.0cm}
    \includegraphics[scale=0.75]{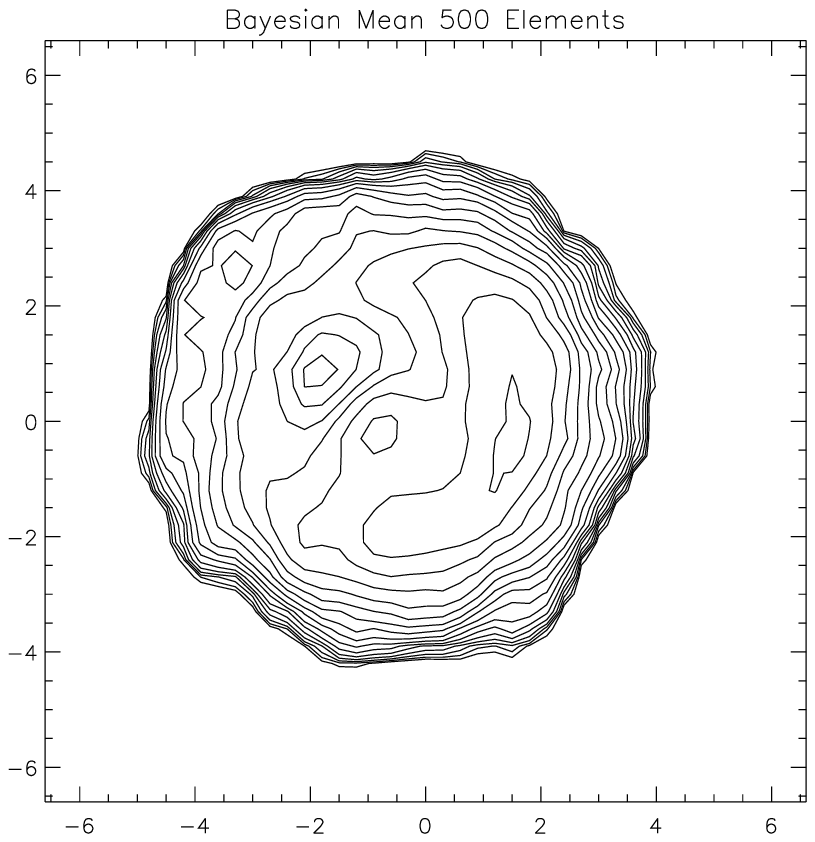}
  \end{minipage}
  \begin{minipage}{10.0cm}
    \includegraphics[scale=0.75]{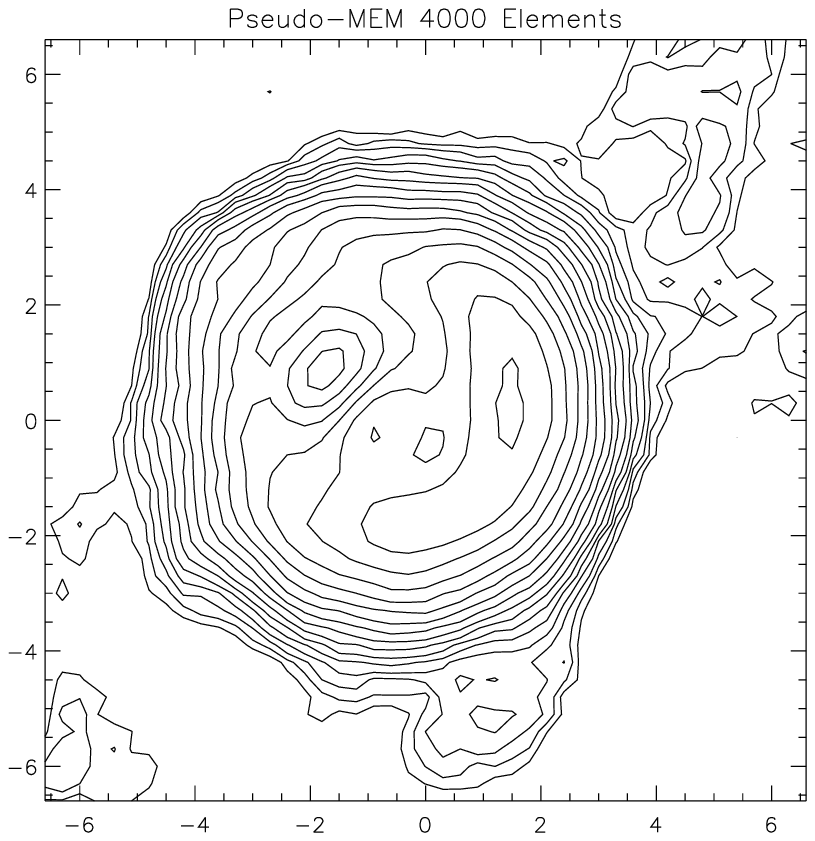}
  \end{minipage}
  \begin{minipage}{7.0cm}
    \includegraphics[scale=0.75]{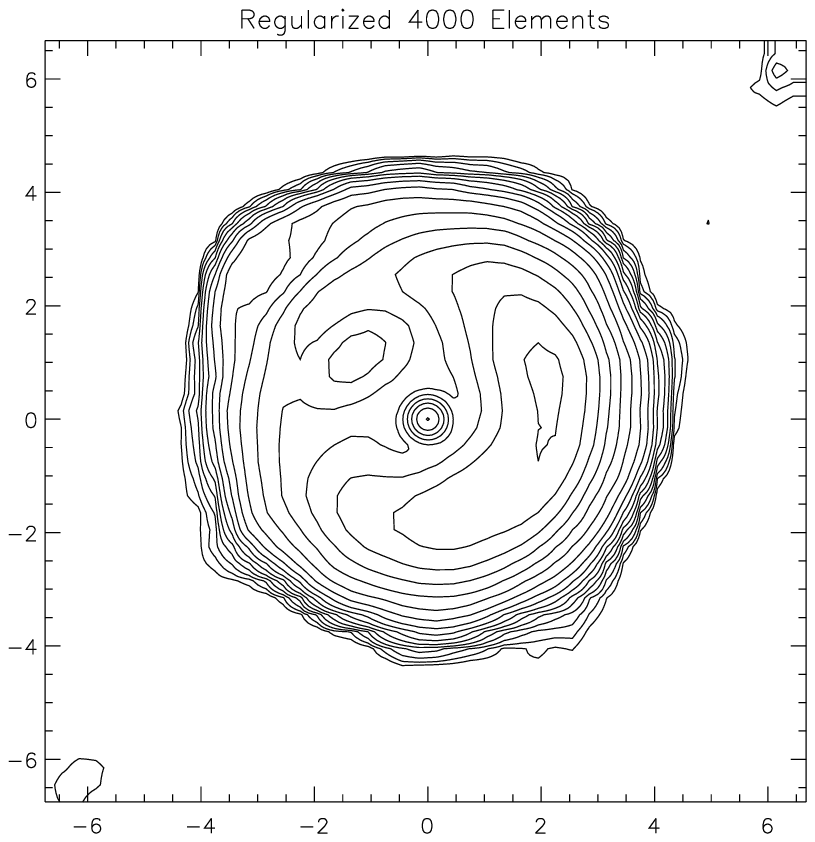}
  \end{minipage}
\vspace{-0.2cm}
\caption{Top row: Model data set 1 from Ref.~\citenum{Lawson04}, with the
  reconstructed image from the `winning' algorithm BSMEM. Middle row:
  Mean images constructed with 2000 and 500 flux elements. The mean
  $\chi^2_r$ of the 9000 images that make up these means are 1.0 and
  0.87 respectively. Bottom row: A `pseudo-MEM' map formed by using
  many (4000) elements, fixing $\chi^2_r = 1$, and allowing the total
  flux to be 1.03 rather than 1.0, and a regularized map that includes
  knowledge of a central point-source (see text). Contours represent
  factors of $\sqrt{2}$ in surface brightness.}
\label{figData1Comp}
\end{figure}

MACIM has been tested on previous beauty contest data
sets, data sets from several aperture masking experiments and data
from the IOTA interferometer. In this section, we will examine the
algorithm's performance on an Imaging Beauty
Contest data set from 2004.

The first 2004 data set was chosen because the signal-to-noise for the
second data set was so high that it was hard to tell apart any two
images that fit the data adequately. Figure~\ref{figData1Comp}
compares the model image and the output from the `winning' algorithm
BSMEM to four output images from MACIM. 

The `pseudo-MEM' MAP is very
similar to the BSMEM map. It was created by choosing the optimal value
of $\alpha$ for the MEM regularizer so that the mode image had a
$\chi^2_r$ of 1. This `mode' image contained a mean of 900 images (10\%
of those saved in memory) to reduce background noise and is a default
output. Note, however, that in order to reproduce
the relatively high background levels in BSMEM, the total image flux had to be
set to 1.03. This is a weakness of standard MEM algorithms that find
their roots in radio interferometry, where the zero-baseline
correlation (i.e. the total flux) is assumed unknown. 

The Bayesian
mean maps with $\lambda= 2000$ and $\lambda = 500$ are both very
similar to the MEM map. However, there is essentially zero background
in the map made with $\lambda=500$, because of the large $\chi^2$
penalty of 1/500th of the flux moving out of the central region,
compared to 1/2000th of the flux moving out of this region. Limiting
the number of flux elements is therefore a powerful regularizer. For
this data set, one could argue that the optimal number of image
elements is about 2000, because with 2000 elements the mean value of
$\chi^2_r$ is 1.0 at unity temperature.

Using the full output of the chain with $\lambda=2000$, we can answer
the question ``Is the feature in the top right hand corner real?'' with
some degree of statistical robustness. Of course, the answer will be
slightly different depending on the value of $\lambda$, so answering
these kinds of questions with MACIM is still work-in-progress. Given
that there are 2000 flux elements, an appropriate question phrasing is
``What is the confidence level for the top right feature containing
more than 1/2000th of the flux''. By adding up the flux elements in a
$3 \times 3$ pixel region for each step in the Markov Chain and
calculating the fraction of time there is non-zero flux, the
confidence level for the feature is only 54\%. 

The possibility of strong regularization with MACIM is also
demonstrated in Figure~\ref{figData1Comp}. The regularized image
is a mean map that includes the dark interaction energy regularizer. The chain began
with the converged output of a previous chain and a point
source with fractional flux 3\% (which could be inferred from
e.g. spectral energy distribution fitting). The point source flux was
left free in the Markov-Chain process, and reached an equilibrium flux
value of $2.6 \pm 0.2$\%. The dark interaction energy (although maybe
not applicable to this image, which has a significant background)
gives the image a sharp edge. The difference between a point-source
model and an image where many flux elements are in the same pixel is
caused by the multiplicity $W$. There is a very small chance that many
(in this case $\sim$100) flux elements can congregate in a single
pixel, so the presence of a point-source becomes strong {\em a priori}
knowledge that influences the final image.

\section{CONCLUSION}

We have demonstrated that the Markov Chain imager MACIM can
reproduce images from {\tt oifits} data with at least equivalent
fidelity to competing algorithms. The main benefit of MACIM are the
simulated annealing algorithm that can converge where self-calibration
does not, and the flexibility in regularization techniques. The source
code, GUI, IDL utilities and compilation instructions for MACIM are
freely available at {\tt http://www.gps.caltech.edu/$\sim$mireland/MACIM}.

\acknowledgments

M.I. would like to acknowledge Michelson Fellowship support from the
Michelson Science Center and the NASA Navigator Program.

\bibliography{../mireland}   %>>>> bibliography data in report.bib
\bibliographystyle{spiebib}   %>>>> makes bibtex use spiebib.bst

\end{document}